\documentclass[12pt]{iopart}
\usepackage{graphicx} 
\usepackage{color}
\usepackage{bm}

\begin{document}

\newcommand \be {\begin{equation}}
\newcommand \ee {\end{equation}}
\newcommand \bea {\begin{eqnarray}}
\newcommand \eea {\end{eqnarray}}

\title[Flux balance and chemical potential]{Influence of flux balance on the generalized chemical potential in mass transport models}

\author{Kirsten Martens$^1$ and Eric Bertin$^2$}

\address{$^1$ Universit\'e de Lyon; Universit\'e Lyon 1, Laboratoire de Physique
de la Mati\`ere Condens\'ee et des Nanostructures; CNRS, UMR 5586,
43 Boulevard du 11 Novembre 1918, F-69622 Villeurbanne Cedex, France\\ 
$^2$ Universit\'e de Lyon, Laboratoire de Physique,
\'Ecole normale sup\'erieure de Lyon, CNRS,
46 all\'ee d'Italie, F-69007 Lyon, France
}
\ead{kirsten.martens@univ-lyon1.fr,eric.bertin@ens-lyon.fr}

\begin{abstract}
In equilibrium systems, the conservation of the number of particles (or mass) leads to the equalization of the chemical potential throughout the
system. Using a non-equilibrium generalization of the notion of
chemical potential, we investigate the influence of disorder
and of the balance of mass fluxes on the generalized chemical potential
in the framework of stochastic mass transport models.
We focus specifically on the issue of local mesurements of the chemical
potential.
We find that while local dynamical disorder does not affect the
measurement process, the presence of large-scale geometrical heterogeneities
(branching geometry) leads to unequal local measurement	 results
in different points of the system.
We interpret these results in terms of mass flux balance,
and argue that the conditions for the global definition of the chemical
potential still hold, but that local measurements fail to capture
the global theoretical value.
\end{abstract}


\section{Introduction}

The question of defining relevant macroscopic control parameters in
non-equilibrium systems still remains largely open
\cite{Jou,Cugliandolo}.
At equilibrium, thermodynamics provides us with intensive parameters
like temperature, pressure and chemical potential, that are uniform
throughout the system even in the presence of heterogeneities,
and that can in most cases be easily measured.
In non-equilibrium steady states, it is natural to try to find
similar types of parameters, and different generalizations of equilibrium
concepts have been proposed
either through statistical approaches often related to entropy notions
\cite{Miller,Chate,Hatano,BDD,Jou06},
or using generalized fluctuation-dissipation relations in theoretical
\cite{CuKuPe,Kurchan,Berthier-Barrat,Puglisi,Crisanti,Ritort,Levine06,MBD-fdr}
and experimental \cite{Danna,Ocio03,Ciliberto,Grenard} contexts.
A few studies \cite{CuKuPe,Barrat,Levine07,Seifert}
have more explicitly considered the question of the equalization
of such parameters throughout inhomogeneous systems,
but this issue has not been settled yet.
Such a question has also been addressed in a previous work
\cite{itp-prl06,itp-pre07}, where the equilibrium definitions
of intensive thermodynamic parameters conjugated to conserved quantities
have been extended to some classes of non-equilibrium
steady-state systems. 
Provided that a condition called ``asymptotic factorization property'' holds
\cite{itp-prl06,itp-pre07}, this approach yields for the out-of-equilibrium
chemical potential $\lambda$ the definition\footnote{Note that this
generalized definition of the chemical potential
differs by a factor $-1/T$ (where $T$ is the temperature) from the
conventional equilibrium definition \cite{balescu}.}
\be \label{lambda-def}
\lambda = \frac{\partial \ln Z}{\partial M},
\ee
where $Z(M)$ is a generalized partition function.
Roughly speaking, the asymptotic factorization condition is satisfied
when only short range correlations are present in the system.
It holds in particular in the absence of correlations, when the joint
probability distribution is factorized (apart from the global
conservation law).

The main properties of this generalized notion of chemical potential
have been studied in \cite{itp-pre07}, where it was shown to exhibit
some interesting equalization properties in some classes
of inhomogeneous non-equilibrium systems. Possible difficulties arising
from the non-equilibrium nature of the systems considered
and especially from the dynamics at the contact between different systems
have also been outlined.

In the present work, we study whether the \emph{globally-defined}
chemical potential given in Eq.~(\ref{lambda-def})
can be evaluated through \emph{local} measurements, and whether measurements
performed on the system at different locations yield the same value,
which is a non-trivial issue when the system is inhomogeneous.
We also investigate how the flux balance
--a strong constraint specific to non-equilibrium situations--
may influence the results of local measurements of the chemical potential.
As simple examples of inhomogeneous non-equilibrium systems,
we consider different models belonging to the class of mass transport
models introduced in Refs.~\cite{Zia04,Zia05}, in which
a globally conserved mass is transferred between neighbouring sites.
We specifically consider both the case of local disorder in which
the dynamics is locally heterogeneous (Section~\ref{sec-ring}),
and the case of a large-scale geometrical heterogeneity related
to the presence of several branches in the system
(Section~\ref{sec-3branches}).

\section{Mass transport model on a ring with local disorder}
\label{sec-ring}

\subsection{Definition of the model}

\begin{figure}[thbp]
\centering\includegraphics[height=4.5cm,clip]{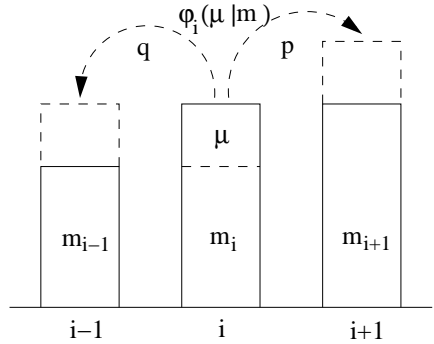}
\caption{Schematic drawing of the dynamics of the mass transport model. A fraction $\mu$ of the mass $m_i$ situated on site $i$ is transferred according to the local conditional rates $\varphi_i(\mu|m_i)$, either to site $i+1$ with probability $p$, or to site
$i-1$ with probability $q=1-p$.}
\label{fig-dynamics}
\end{figure}

We focus here on a one-dimensional mass transport model with
periodic boundary conditions, that is on a ring geometry.
On each site $i=1,\ldots,N$ resides a positive mass $m_i$.
The continuous time stochastic dynamics, which
preserves the total mass $M=\sum_{i=1}^N m_i$, is defined as follows
(see Fig.~\ref{fig-dynamics}).
An amount of mass $\mu$ is transferred from site $i$, containing the mass $m_i$,
to site $i+1$ with a probability per unit time $p\,\varphi_i(\mu|m_i)$,
and to site $i-1$ with a probability per unit time $q\,\varphi_i(\mu|m_i)$,
where $q=1-p$ (note that $N \equiv 0$ and $N+1 \equiv 1$ due to periodic
boundary conditions). The rate $\varphi_i(\mu|m_i)$ is defined as
\begin{equation}\label{mtm-rate}
\varphi_i(\mu|m_i) = v(\mu)\, \frac{f_i(m_i-\mu)}{f_i(m_i)}\;,
\end{equation}
where $v(\mu)$ and $f_i(m)$ are arbitrary positive functions,
with $f_i(m)$ possibly site-dependent.
Thus transport is biased (except if $p=q$), which generates a flux of mass
along the ring.
With the above rate, the steady-state distribution takes the form
\cite{Zia04} 
\begin{equation} \label{dist-mtm}
P(\{m_i\})=\frac{1}{Z(M)} \prod_{i=1}^N f_i(m_i) \,
\delta\left(\sum_{i=1}^N m_i-M\right)\;,
\end{equation}
with $\delta(x)$ the Dirac delta function, and where
the partition function $Z(M)$ is defined as
\begin{equation}
Z(M)= \int \prod_{i=1}^N [dm_i\, f(m_i)]\, \delta\left(\sum_{i=1}^N m_i-M\right)\; .
\end{equation}
Note that the function $v(\mu)$ does not influence
the steady-state distribution, but only the dynamics.
When performing numerical simulations, we use throughout the paper
$v(\mu)=1$.

\subsection{Globally-defined chemical potential.}

Let us here consider for $f_i(m_i)$ the simple form $f_i(m_i)=m_i^{\eta_i-1}$
with $\eta_i>0$ for all $i$.
To calculate the generalized chemical potential corresponding
to the conserved mass in the system
we need to find the dependence of the partition function $Z$ on $M$:
\begin{equation}
Z(M)= \int \prod_{i=1}^N [dm_i\, m_i^{\eta_i-1}]\, \delta\left(\sum_{i=1}^N m_i-M\right)\; ,
\end{equation}
where the integrals are over the positive real axis.
A simple rescaling $m_i=x_i M$ reveals the searched dependence:
\begin{eqnarray}
Z(M)&=&M^{\sum_{i=1}^N \eta_i-1} \int \prod_{i=1}^N [dx_i\, x_i^{\eta_i-1}]\, \delta\left(\sum_{i=1}^Nx_i-1\right)\nonumber\\
&=& D_N M^{N\overline{\eta}-1}
\end{eqnarray}
with $\overline{\eta}=N^{-1}\sum_{i=1}^N \eta_i$, and where
$D_N$ is a constant independent of M. 
The generalized chemical potential is obtained from the derivative of $\ln Z$
\be
\lambda = \frac{d \ln Z}{d M}
= \frac{N\overline{\eta} -1}{M}
\ee
leading in the thermodynamic limit $N \to \infty$ to
\be \label{eq-itp0}
\lambda = \frac{\overline{\eta}}{\rho}\;,
\ee
where $\rho=M/N$ denotes the average density.

\subsection{Local measurement of the chemical potential.}
\label{loc-measure-chem}

Once the chemical potential has been theoretically defined, an important
issue is to know whether it can be measured. One possible way
to perform a measurement is to connect to the considered system
a probe system, with a much smaller size in order not to significantly
perturb the main system. The probe system is assumed to have
a known equation of state, so that its chemical potential
can be deduced from its average mass. The connection between the two systems
is in general local --just like when a thermometer is put into contact
with a system to measure its temperature.
As a result, it is important to verify that the measurement result is the
same wherever the connection is made. This issue becomes non-trivial
when the system is inhomogeneous.

\begin{figure}[htpb]
\centering\includegraphics[height=4cm,clip]{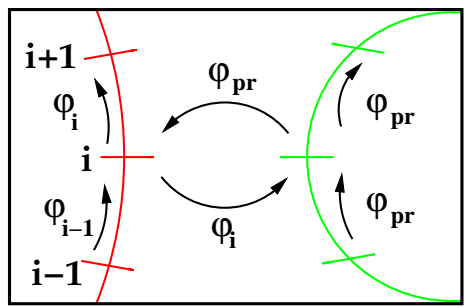}
\caption{Sketch of the contact of a small probe system (right) with a large system (left). Mass is transferred in the direction of the arrows with the indicated transport rates (see text).}
\label{fig-itp-contact}
\end{figure}

To check this feature, we consider the above inhomogeneous mass transport model
on a ring with transport rates $\varphi_i(\mu|m_i)$ defined
according to Eq.~(\ref{mtm-rate}) with $f_i(m)=m^{\eta_i-1}$.
We choose a sinusoidal space dependence for $\eta_i$, of the form
$\eta_i=2+\sin(2\pi i/N)$, $i=1,\ldots, N$.
We also set $p=1$ in the numerical simulations.
As a probe system, we use a homogeneous mass transport model,
with site-independent rates $\varphi_\mathrm{pr}(\mu|m_i)$ 
defined with $f_{\mathrm{pr}}(m)\equiv 1$, that we successively attach
to sites $i=N/4$ and $i=3N/4$ of the inhomogeneous system
(see Fig.~\ref{fig-itp-contact}).

\begin{figure}[htbp]
\centering\includegraphics[height=5.5cm,clip]{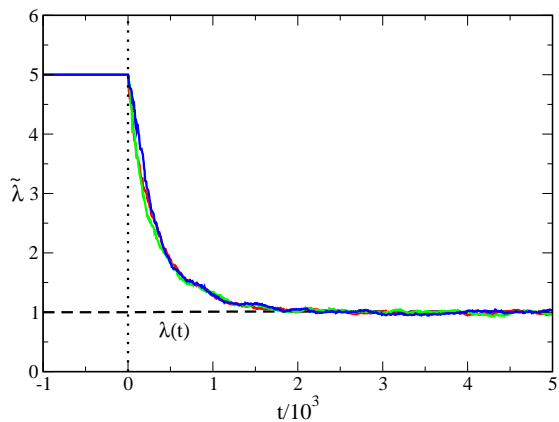}
\caption{(Color online) Chemical potentials $\tilde{\lambda}(t)$ of the probe system (upper curves)
and $\lambda(t)$ of the main system (lower curve) during the measurement process,
in three different situations: (i) Inhomogeneous system with
$\eta_i=2+\sin(2\pi i/N)$, and probe connected to $i=N/4$ (red curve);
(ii) Same system, with the probe connected to $i=3N/4$ (green curve);
(iii) Inhomogeneous system with random and uncorrelated values
of $\eta_i$ on each site, drawn from a uniform distribution on
the interval $1<\eta_i<3$ (blue curve).
In all cases, the mean value $\overline{\eta}=2$ is the same.
The contact is switched on at time $t=0$ (vertical dotted line).
Initial densities are $\rho=2$ for the main system and $\rho_{\mathrm{pr}}=0.2$
for the probe. All measurements converge to the theoretically expected value
(dashed line).
System size $N=15870$, probe size $N_{\mathrm{pr}}=512$.}
\label{fig-itp-meter-inhom}
\end{figure}

Fig.~\ref{fig-itp-meter-inhom} shows that the chemical potential $\tilde{\lambda}(t)$ 
of the probe converges to the chemical potential of the main system
(computed from its equation of state).
The results are seen to be independent of the location
where the probe is attached to the inhomogeneous system.
We further observe that even in a completely disordered system,
in which the $\eta_i$'s are independent and identically distributed
quenched random variables, the measured
value of the chemical potential is still the same as long as the mean value
$\overline{\eta}=1/N\sum_i\eta_i$ remains the same as in the deterministic
case.

\subsection{Relation with the balance of mass flux.}

We have seen that, in this simple model with a ring geometry,
the locally-measured chemical potential coincides with the global one,
in spite of the heterogeneities which makes the local density site-dependent.
This result can be interpreted from the one-site probability distribution
$p_i(m_i)$.
Considering the rest of the system as a reservoir of mass, one finds
for the probability distribution on site $i$
\be \label{dist-pi}
p_i(m_i) = \frac{1}{Q_i} \, f_i(m)\, e^{-\lambda m_i},
\ee
where $Q_i$ is a normalization constant.
Hence the local chemical potential is indeed the same everywhere in the
system.

Interestingly, this result can also be given an alternative interpretation
in terms of flux balance.
Starting from Eq.~(\ref{dist-pi}), one can relate the flux and the chemical
potential $\lambda$. The total mass flux $\Phi_i$ crossing site $i$
can be expressed as
\be
\Phi_i = (p-q)\int_0^{\infty} dm\, p_i(m) \int_0^m d\mu\, \mu\, \varphi_i(\mu|m).
\ee
which can be rewritten, using Eqs.~(\ref{dist-pi}) and (\ref{mtm-rate}), as
\be \label{eq-phi-lambda}
\Phi_i = (p-q)\int_0^{\infty} d\mu\, \mu v(\mu)\, e^{-\lambda \mu}.
\ee
One thus obtains that $\Phi_i=\Phi$ is independent of the site considered,
as expected from the steady-state flux balance in this linear geometry.
Alternatively, one could interpret Eq.~(\ref{eq-phi-lambda}) by saying that
the uniformity of the local chemical potential results from the flux balance.
Hence one can guess that flux balance plays an important role in
the determination of the local chemical potential.
This role will appear even more clearly in the example considered
in the next section, where the geometry is no longer purely linear.

\section{Model with three branches}
\label{sec-3branches}

\subsection{Definition of the model}
\label{sec-def-3branch}

\begin{figure}[thbp]
\begin{minipage}[b]{0.7\linewidth}
\centering\includegraphics[width=\linewidth,clip]{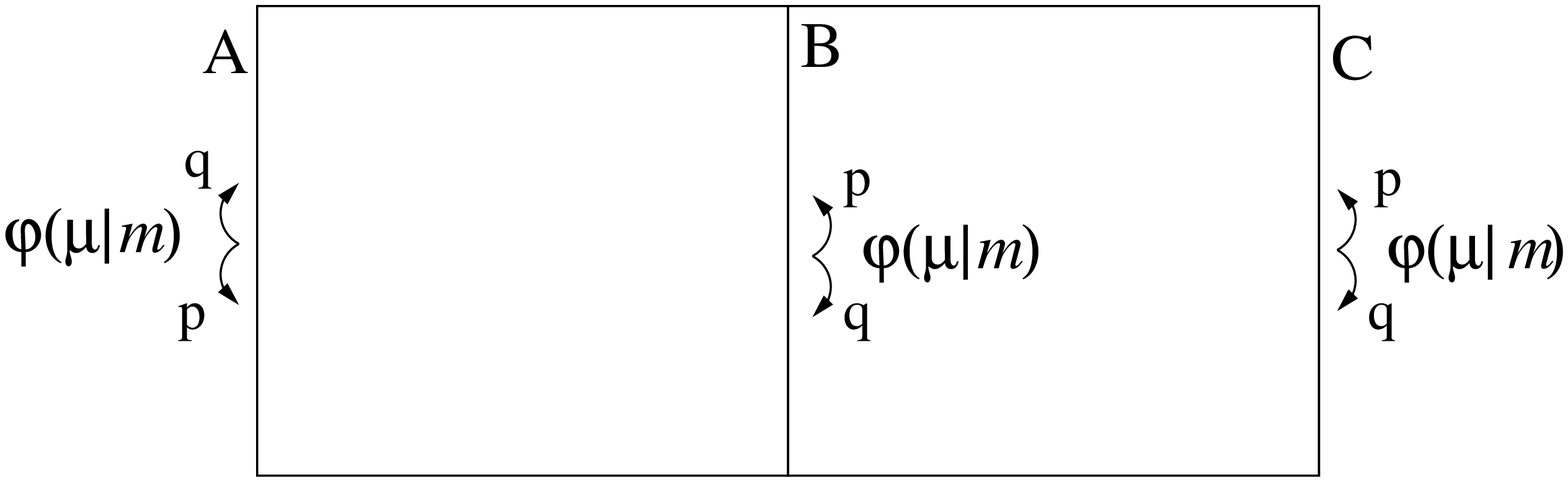}
\vspace{0.5cm}
\end{minipage}
\begin{minipage}[b]{0.3\linewidth}
\centering\includegraphics[width=\linewidth,clip]{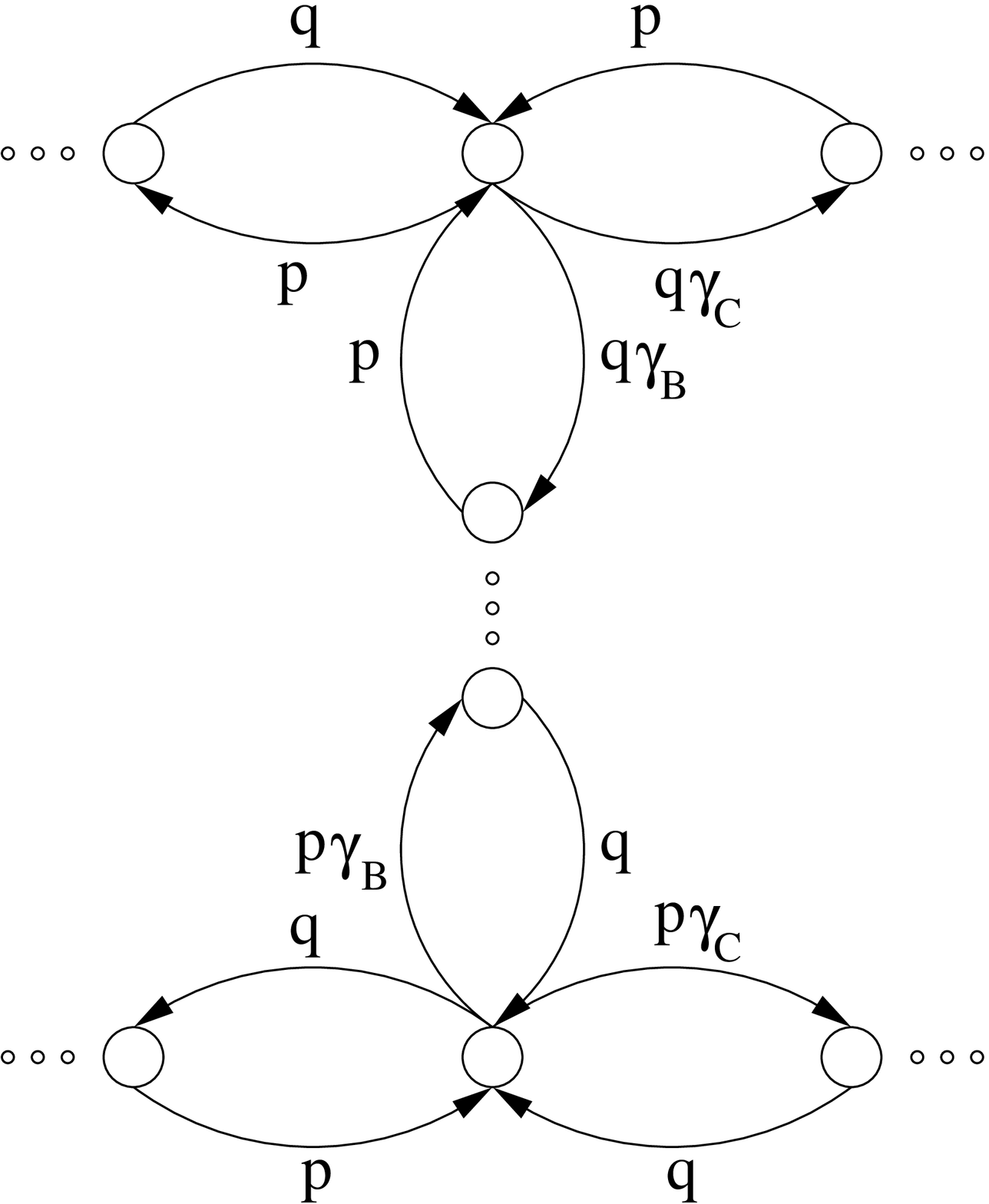}
\end{minipage}
\caption{Sketch of the model with three branches. Left panel: Mass is transported along each oriented branch according to the rate $p\,\varphi(\mu,m)$ in the positive direction, and to the rate $q\,\varphi(\mu,m)$ in the negative direction.
Right panel: Zoom on the branching points, where specific rules are taken into account,
some of the transfer rates being reweighted by factors $\gamma_B$ and $\gamma_C$.}
\label{fig-branches}
\end{figure}

We now investigate the issue of the local measurement of the chemical
potential in a model where the local dynamics is essentially homogeneous,
but where the heterogeneity results from a branching geometry.
We consider a mass transport model with three branches, corresponding to the geometry displayed in Fig.~\ref{fig-branches}.
The three branches are assumed to be oriented.
The transfer rate from site $i$ to the neighbouring
site is $p\, \varphi(\mu|m_i)$ along the positive direction (according to the
orientation of each branch) and $q\, \varphi(\mu|m_i)$ along the negative direction.
The rate $\varphi(\mu|m)$ is defined according to Eq.~(\ref{mtm-rate}),
with a site-independent weight function $f(m)$.
At the branching points ($A \to B$ or $C$), probability rates for the transfer
to branches $B$ and $C$ are reweighted by factors
$\gamma_B$ and $\gamma_C$, as shown on Fig.~\ref{fig-branches}.
For instance, at the lower branching point on Fig.~\ref{fig-branches},
the transfer from branch $A$ occurs with rate
$\gamma_B p\, \varphi(\mu|m_i)$ to branch B, and with rate
$\gamma_C p\, \varphi(\mu|m_i)$ to branch C.
Due to this specific geometry, the probability distribution does not necessarily factorize, even with the choice of transport rates $\varphi(\mu|m)$ given in Eq.~(\ref{mtm-rate}).

For mass transport models defined on an arbitrary graph,
a sufficient condition for the factorization of the probability
distribution has been given in Ref.~\cite{graph-mtm}.
In the most general case, such models are defined by transport rates
$\varphi_{ij}(\mu|m_i)$ from site $i$ to site $j$ of the form
\be \label{mtm-rate-graph}
\varphi_{ij}(\mu|m_i) = v_{ij}(\mu)\, \frac{f_i(m_i-\mu)}{f_i(m_i)}
\ee
where the function $v_{ij}(\mu)$ is identically zero if there is no directed link from
$i$ to $j$. If the condition 
\be \label{structure}
\sum_{j (\ne i)} v_{ij}(\mu) = \sum_{j (\ne i)} v_{ji}(\mu)
\ee
holds for every site $i$, the probability distribution factorizes, with the local
probability weight given by $f_i(m_i)$ \cite{graph-mtm}, as in Eq.~(\ref{dist-mtm}).

In the case of the model with three branches, the branching sites violate the sufficient condition for factorization given Eq.~(\ref{structure}), which is a strong indication that strict factorization does not hold.
Hence the exact solution of this model is not known, and it is not clear
a priori whether the present model satisfies or not the
``asymptotic factorization condition'', which is a key criterion
for the existence of a globally defined chemical potential.
Indeed, although the strict factorization property is likely to be violated,
one might ask whether this violation is 'localized' around the branching
points and if the local chemical potential in the bulk of the branches remains uniform.
It is thus interesting in this situation to perform
numerical measurements of the local chemical potential.

\subsection{Measure of the chemical potential with a probe system}
\label{sec-3branch-probe}

\begin{figure}[thbp]
\centering\includegraphics[height=5.5cm,clip,]{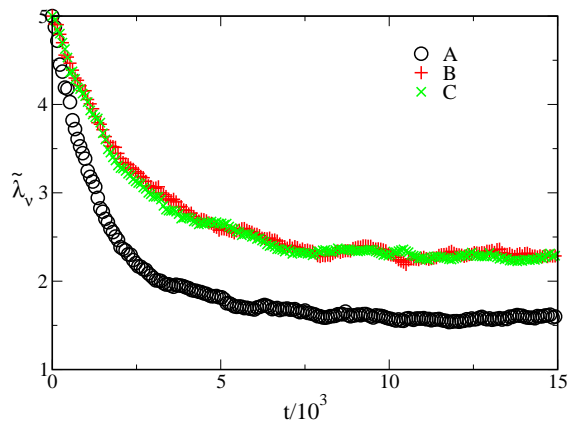}
\caption{Chemical potentials $\tilde{\lambda}_\nu$, $\nu=A,B,$ or $C$ measured with a probe system attached to branches $A, B$ and $C$,
plotted as a function of time $t$ after the contact is switched on at $t=0$.
The transport rates are the same in all branches and in the probe system,
 $\varphi(\mu|m_i)\equiv 1$.
The dynamics at the contact is defined  as shown in Fig.~\ref{fig-itp-contact}.
The initial value of the global density is $\rho=0.5$ in the main system and
$\rho_{\mathrm{pr}}=0.2$ in the probe system.
All three branches have the same size $N_b = 43348$, and the probe size
is $N_\mathrm{pr}=1024$.
Measurements on each branch have been performed during separate runs.}
\label{fig-abc}
\end{figure}

To test this issue, we measure the chemical potential with a probe system attached to a bulk site of the branch considered,
in analogy to the procedure explained in Sect.~\ref{loc-measure-chem}.
The transport rates are the same in all branches as well as in the probe system, $\varphi(\mu|m)\equiv1$.
All three branches have the same number of sites $N_b$, so that
the system size is $N=3N_b$ (to be specific, the two branching points
are included in branch A). The branching points contribute to the number of sites
in branch $A$. Simulations are done with $p=1$ and $\gamma_B=\gamma_C=\frac{1}{2}$.
The results of the numerical implementation of the measurement are shown
in Fig.~\ref{fig-abc}. The locally measured chemical potential is denoted as
$\tilde{\lambda}_\nu$ in branch $\nu=A$, $B$, $C$.
It turns out that we obtain equal values for the locally measured chemical potential
in branches $B$ and $C$, but a very different value for branch $A$.

\subsection{Measure of the chemical potential through local fluctuations}
\label{sec-3branch-fluct}

To check whether this result obtained by probing the system depends on the measurement method, we apply in the following an alternative measurement technique that does not require any external device.
The idea, introduced in an earlier work \cite{itp-pre07},
is based on a direct measurement of the fluctuations
in small subsystems of different sizes, within each branch.
We briefly sketch the procedure in the following.
Let us first define the quantity $g_{\nu}$ ($\nu=A, B$, or $C$) as the variance
of the total mass $M_{\nu}$ in branch $\nu$ divided by the number of sites
$N_{\nu}$ in this branch:
\be
g_\nu = \frac{\langle M_{\nu}^2\rangle-\langle M_{\nu}\rangle^2}{N_\nu}\;.
\ee
We have checked numerically that $g_\nu$ does not depend on the size of the subsystem
chosen within a given branch. Consequently $g_\nu$ is an intensive quantity depending only on the local density, $g_\nu=g_\nu(\rho_\nu)$. For a detailed description of this procedure see \cite{itp-pre07}.

\begin{figure}[htbp]
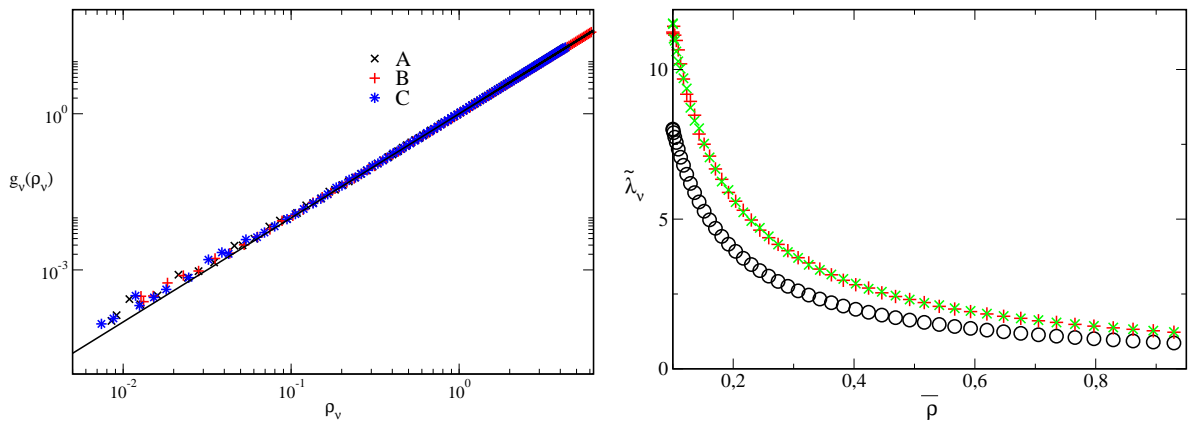

\centering\includegraphics[height=5.5cm,clip,]{figure6a.eps}
\hfill
\centering\includegraphics[height=5.5cm,clip,]{figure6b.eps}
\caption{Left panel: measure of the function $g_\nu(\rho_{\nu})$ in each branch $A$, $B$ and $C$. The same functional form, very close to the
theoretical form $g(\rho)=\rho^2$ corresponding to a homogeneous system,
is obtained in each case.
Transport rates: $\varphi(\mu|m)\equiv 1$. All branches have the same size $N_b=341$.
Right panel: locally-measured chemical potential
$\tilde{\lambda}_{\nu}=\Lambda(\rho_\nu)$ in each branch
$\nu = A$, $B$, and $C$, plotted as a function of the {\it total} density in the system
$\overline{\rho}=M/N$ (from simulations).
The expected value of $\tilde{\lambda}_\nu$ for $\varphi(\mu|m)=1$
is $\tilde{\lambda}_\nu=\rho_\nu^{-1}$ (see text).}
\label{fig-variance}
\end{figure}

We choose the same parameters for the dynamics on the three branches as
in Sec.~\ref{sec-3branch-probe}.
Each branch is of the same size $N_b$ and the transport rates are given by $\varphi(\mu|m) \equiv 1$ everywhere. The result of a numerical implementation
of the measurement of the function $g_\nu(\rho_\nu)$ in the three-branch
model is shown in the left panel of Fig.~\ref{fig-variance}.
The functional behaviour of $g_\nu(\rho_\nu)$ is seen to be the same
for the three branches, and we denote this function simply as
$g(\rho_\nu)$. We further observe on Fig.~\ref{fig-variance}
that the numerically measured $g(\rho_\nu)$ is very close to the theoretical
value $g(\rho_\nu)=\rho^2$ corresponding to a homogeneous system with the same
local dynamics $\varphi(\mu|m)=1$, so that we shall use this theoretical expression
in the following.

Using the grandcanonical ensemble \cite{itp-pre07},
the chemical potential can be related to the function $g(\rho_\nu)$ according to
\be
g(\rho_\nu)=-\frac{d\rho_\nu}{d\tilde{\lambda}_\nu} \;,
\ee
from which the relation
\be
\tilde{\lambda}_\nu = \Lambda (\rho_\nu) \equiv \int_{\rho_{\nu}}^{\infty} \frac{d\rho}{g(\rho)}
\ee
follows.
This means that the functional behaviour of the local chemical potential $\Lambda$ with density $\rho_\nu$, that is the local equation of state, is the same in the three branches, namely $\Lambda(\rho_\nu)=\rho_\nu^{-1}$ for the specific dynamics chosen here.
This result is not surprising since the local dynamics is the same everywhere
and although the probability distribution does not factorize, only weak
correlations are expected in the bulk of each branch.
Taking into account the fact that the local densities $\rho_\nu$ in the three branches
are not equal, it follows that the values $\tilde{\lambda}_\nu=\Lambda(\rho_\nu)$
of the local chemical potential of the three branches
differ as well: $\tilde{\lambda}_A\neq\tilde{\lambda}_B=\tilde{\lambda}_C$.

\begin{figure}[htbp]
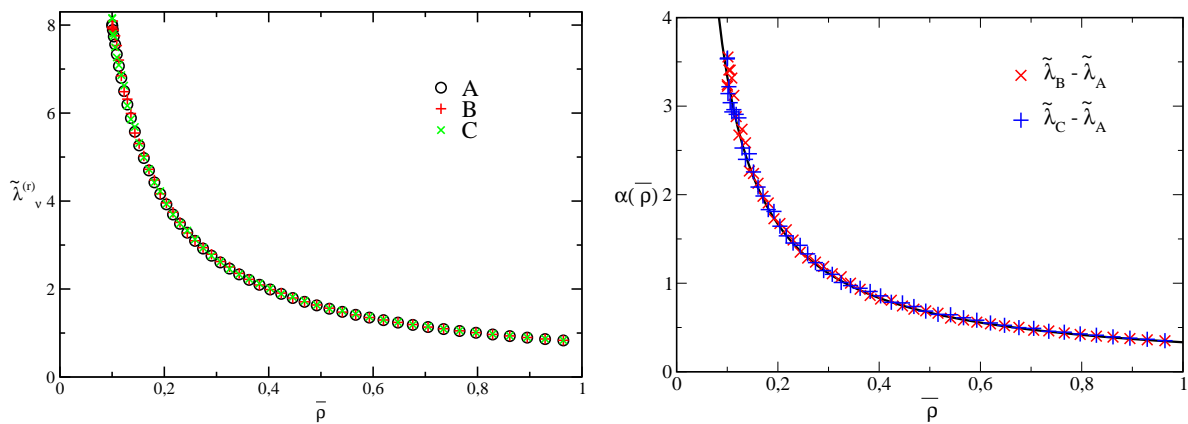

\centering\includegraphics[height=5.5cm,clip,]{figure7a.eps}
\hfill
\centering\includegraphics[height=5.5cm,clip,]{figure7b.eps}
\caption{Left panel:  rescaled values $\tilde{\lambda}^{(r)}$, defined as
$\tilde{\lambda}_A^{(r)}=\tilde{\lambda}_A$ and $\tilde{\lambda}_{B,C}^{(r)}=\tilde{\lambda}_{B,C}/\sqrt{2}$, plotted as a function of $\overline{\rho}$, showing a collapse of the data (same data as in the right panel of Fig.~\ref{fig-variance}).
Right panel: differences in the locally-measured chemical potential of system $B$ and $A$ ({\color{red} $\times$}), and system $C$ and $A$ ({\color{blue} +}) as a function of $\overline{\rho}$
compared to the analytical prediction $\alpha(\overline{\rho})=1/(3\overline{\rho}$ (black line) as discussed in Sect.~\ref{consistency}.}
\label{fig-collapse}
\end{figure}

Though this discrepancy points to a weakness of the concept
of non-equilibrium chemical potential, it can however be understood
by taking into account the balance of fluxes.
Starting from the expression (\ref{eq-phi-lambda}) of the flux $\Phi$,
we first note that if the function $v(\mu)$ in the transport rates
is identically 1, $v(\mu)\equiv1$, then the flux is given by $\Phi=1/\tilde{\lambda}^2$
(note that the expression of the flux is independent of the form of $f(m)$).
We use here the local chemical potential $\tilde{\lambda}$, as the flux is governed
by the local dynamics --see Eq.~(\ref{eq-phi-lambda}). We shall come back to this point
in Sec.~\ref{sec-3branch-ZRP}.
In the bulk of the branches, where the correlations are expected to be small,
the different fluxes are given by
$\Phi_A=1/\tilde{\lambda}_A^2$ and $\Phi_C=\Phi_B=1/\tilde{\lambda}_B^2$
(fluxes in branches $B$ and $C$ are equal since $\gamma_B=\gamma_C$).
Moreover we know that the flux in branch $A$ is twice the flux in branch $B$, which leads to the following relation between the chemical potentials of the different branches:
\begin{equation} \label{lambdaAB1}
\tilde{\lambda}_A=\frac{\tilde{\lambda}_B}{\sqrt{2}}=\frac{\tilde{\lambda}_C}{\sqrt{2}}\;.
\end{equation}
This relation has been verified numerically for $\varphi(\mu|m)\equiv 1$, where we expect analytically $\tilde{\lambda}_\nu=\rho_\nu^{-1}$, $\nu=A,B,C$ -- see the right panel of Fig.~\ref{fig-variance} and left panel of Fig.~\ref{fig-collapse} for the numerical results.
We can generalize this flux balance argument
to situations where branches $B$ and $C$ do not have the same flux,
in which case the relation reads:
\be
\tilde{\lambda}_A = \sqrt{\frac{\Phi_B}{\Phi_A}}\, \tilde{\lambda}_B = \sqrt{\frac{\Phi_C}{\Phi_A}}\, \tilde{\lambda}_C\;.
\label{lambdaAB2}
\ee
Note that at equilibrium, when the fluxes vanish, this correction is not
present and all chemical potentials equalize.
But as soon as a little bias is introduced in the dynamics,
the value of the chemical potential in branch $A$ differs from that of the
two other branches. Let us emphasize that this difference is not
perturbative with respect to the bias: $\tilde{\lambda}_A - \tilde{\lambda}_B$ does
not go to zero when $p-q \to 0$, but rather remains constant
as long as $p > q$.

From a more theoretical perspective, the discrepancy between
$\tilde{\lambda}_A$, $\tilde{\lambda}_B$ and $\tilde{\lambda}_C$ questions the validity
of the asymptotic factorization condition \cite{itp-prl06,itp-pre07}
required to define a global chemical potential from Eq.~(\ref{lambda-def}).
It would thus be interesting to verify explicitly whether this
asymptotic factorization property holds or not.
This is a difficult task because (to our knowlegde) the exact joint probability
distribution is not known for generic rates in the present geometry,
as it does not fulfill Eq.~(\ref{structure}).
Yet, the specific case of the Zero Range Process (ZRP),
where masses are discrete, turns out to be solvable
and thus deserves to be investigated in more details.

\subsection{A solvable case with discrete masses}
\label{sec-3branch-ZRP}

The ZRP case, which has been intensively studied in the literature
\cite{Evans-Rev05}, corresponds to the choice $v(\mu)=\delta(\mu-1)$,
so that masses $m_i$ take integer values, denoted as $n_i$ in the following.
We study here the ZRP in the three branch geometry illustrated
on Fig.~\ref{fig-branches}.
The dynamics is defined as in Sec.~\ref{sec-def-3branch}, except that
$\mu$ can only take the value $\mu=1$.
To simplify the calculations, we set $\gamma_B=\gamma_C=1$ and choose $p>q$.

The relation (\ref{eq-phi-lambda}) between flux and local chemical potential
reads in this case $\Phi_\nu=e^{-\tilde{\lambda}_\nu}$.
Given that fluxes are different in the three branches, it is clear that the
chemical potentials are not equal.
More quantitatively, the balance of fluxes $\Phi_A=\Phi_B+\Phi_C$ implies
$\tilde{\lambda}_A=-\ln(\Phi_A /\Phi_B)+\tilde{\lambda}_B$.
From this expression we see that there is now a shift, given by the logarithm
of the ratio of the fluxes, in the value of the chemical potential.

To better understand the origin of this shift, we compute the steady-state
distribution of the ZRP with three branches.
Interestingly, it has been shown that the steady-state distribution of the ZRP
on an arbitrary graph remains factorized \cite{graph-mtm}.
Slightly rephrasing the results of Ref.~\cite{graph-mtm},
one finds that the distribution $P(\{n_i\})$ is given by
\be
P(\{n_i\} = \frac{1}{Z(M)}\, \left( \prod_{i=1}^N f(n_i)\, z_i^{n_i} \right)
\, \delta_{\sum_i n_i,M}
\ee
where $\delta$ is the Kronecker delta symbol, and where
the local ``fugacities'' $z_i$ satisfy for all $i$ the equation
\be \label{balance-yji}
\sum_{j (\ne i)} v_{ij}\, z_i = \sum_{j (\ne i)} v_{ji}\, z_j
\ee
with $v_{ij} \equiv v_{ij}(1)$ defined in Eq.~(\ref{mtm-rate-graph}).
In the present three-branch model, $v_{ij}$ is equal to $p$ if there is a positively
oriented link from $i$ to $j$, and to $q$ if the link is negatively
oriented; $v_{ij}=0$ in the absence of link.
Note that the fugacities $z_j$ are defined only up to an overall
arbitrary factor.
In the present model, condition (\ref{balance-yji}) reads, for all site $i$
different from a branching point:
\be
p z_{i-1} - z_i + q z_{i+1} = 0.
\ee
The general solution of this linear equation (valid separately on each branch)
is a linear combination of solutions of the form $z_j=r^j$, with a
parameter $r$ obeying the relation
\be \label{eq-r}
q r^2 -r+p=0.
\ee
Eq.~(\ref{eq-r}) has two solutions, $r_1=1$ and $r_2=p/q>1$.
As a result, $z_j$ can be expressed on each branch as a linear combination
of the two independent solutions $r_1^j$ and $r_2^j$,
\be \label{zj-zK}
z_j=\overline{z}_{\nu}\left[1 + K_{\nu} \left(\frac{q}{p}\right)^{N_\nu-j}
\right]
\ee
where $\overline{z}_{\nu}$ and $K_{\nu}$ are constants ($\nu=A$, $B$, or $C$) and
$N_\nu$ denotes the number of sites in branch $\nu$. The total number of sites in
the system is given by $N=N_A+N_B+N_C+2$ (the last term accounts for the two nodes).
Taking into account Eq.~(\ref{balance-yji}), formulated for the two
branching points, one can match the expressions (\ref{zj-zK}) corresponding to
different branches and determine the constants $K_{\nu}$
as well as the ratios $\overline{z}_A/\overline{z}_B$
and $\overline{z}_A/\overline{z}_C$.
In the limit where the sizes $N_A$, $N_B$ and $N_C$
of the three branches go to infinity, one finds
$\overline{z}_A/\overline{z}_B=\overline{z}_A/\overline{z}_C=2$.
As $z_j$ is defined up to an overall prefactor, one can choose for instance
$\overline{z}_B=\overline{z}_C=1$ and $\overline{z}_A=2$.
Given that the terms proportional to $K_{\nu}$ are exponentially decaying
corrections, it turns out that $z_j$ is almost constant, and equal to
$\overline{z}_\nu$, on each branch $\nu$.

Hence the definition of the chemical potential proposed in 
\cite{itp-prl06,itp-pre07} can indeed be applied.
Splitting the system into two parts, branch $A$ on one side, and branches $B$ and $C$
on the other side, one obtains
\be
\lambda_A = \frac{\partial \ln Z_A}{\partial M_A},
\qquad
\lambda_{BC} = \frac{\partial \ln Z_{BC}}{\partial M_{BC}}.
\ee
with $M_A=\sum_{i \in A} n_i$ and $M_{BC} = \sum_{i \in B,C} n_i$ the respective
masses of the two subsystems.
Since $\overline{z}_A=2$, we get, neglecting the exponential corrections appearing in
Eq.~(\ref{zj-zK})
\be \label{eq-ZAZB-ZRP}
Z_A = 2^{M_A} Z_A^{(0)}, \qquad Z_{BC} = Z_{BC}^{(0)},
\ee
where $Z_A^{(0)}$ and $Z_{BC}^{(0)}$ are the standard
partition function of these two subsystems, that would be obtained by
taking $z_i=1$ for all $i$.
This results, in the large size limit, in
\be \label{eq-lambda-ZRP}
\lambda_A = \ln 2 + \tilde{\lambda}_A, \qquad
\lambda_{BC} = \tilde{\lambda}_{BC}.
\ee
We thus recover in this way the result directly obtained from
the balance of fluxes, showing that this result is actually consistent with the
asymptotic factorization condition
on which the definition of the chemical potential is built
(let us recall that the probability distribution of the present ZRP model is factorized).
Note that the shift $\ln 2$ is independent
of the precise value of the bias $p-q$, as long as this bias
is non-zero. If $p=q$, equilibrium is recovered, and the bias vanishes.
The presence of this bias is thus a genuine non-equilibrium effect,
which appears non-perturbatively, in a discontinuous way.

It is also interesting to note that the shift $\ln 2$ cancels out from the one-site
mass distribution.
Indeed, one has for a site $i$ in the bulk of branch $A$, using Eq.~(\ref{eq-lambda-ZRP})
\be
p_i(n) = \frac{1}{Q_A}\, f(n)\, 2^n e^{-\lambda_A n}
= \frac{1}{Q_A}\, f(n)\, e^{-\tilde{\lambda}_A n}
\ee
and for a site in branch $B$ or $C$
\be
p_i(n) = \frac{1}{Q_{BC}}\, f(n)\, e^{-\tilde{\lambda}_{BC} n}.
\ee
Hence both distributions have exactly the same form as a function of the local
chemical potential, making the shift undetectable from a local measurement.

\subsection{Consistency between flux balance and equality of the chemical potentials}
\label{consistency}

We have seen in the ZRP case that the discrepancy between the local chemical potentials
$\tilde{\lambda}_\nu$ can be explained by the presence of exponential factors
that differ from one subsystem to the other.
It is tempting to try to generalize this scenario to the case
of the mass transport model with continuous masses, although no analytical
solution is available in this case. At a heuristic level,
a more general form of Eq.~(\ref{eq-ZAZB-ZRP}) can be proposed, namely
\be \label{eq-ZAZB-gen}
Z_A = e^{\alpha(\overline{\rho}) M_A} Z_A^{(0)}, \qquad Z_{BC} = Z_{BC}^{(0)},
\ee
since the global average density $\overline{\rho}$ is fixed, and
can thus a priori enter as a parameter in the exponential factor.
Eq.~(\ref{eq-lambda-ZRP}) is then replaced by
\be \label{eq-lambda-gen}
\lambda_A = \alpha(\overline{\rho}) + \tilde{\lambda}_A, \qquad
\lambda_{BC} = \tilde{\lambda}_{BC}.
\ee
Assuming that the asymptotic factorization condition holds, one
has $\lambda_A = \lambda_{BC}$, so that
\be \label{eq-alpha-rho}
\alpha(\overline{\rho}) + \tilde{\lambda}_A = \tilde{\lambda}_{BC}.
\ee
The constant $\alpha(\overline{\rho})$ is determined as follows.
Given a value of $\overline{\rho}$, one looks for the densities
$\rho_A(\overline{\rho})$ and $\rho_{BC}(\overline{\rho})$ 
satisfying the contraints
\bea
\tilde{\Phi}(\rho_A) &=& 2 \tilde{\Phi}(\rho_{BC}),\\
\overline{\rho} &=& \kappa \rho_A + (1-\kappa)\rho_{BC}
\eea
with $\kappa=N_A/(N_A+N_{BC})$, and where $\tilde{\Phi}(\rho)$ is the value
of the local flux for a local density $\rho$.
The parameter $\alpha(\overline{\rho})$ is then obtained, consistently
with Eq.~(\ref{eq-alpha-rho}), as
\be \label{eq-alpha-rho2}
\alpha(\overline{\rho}) = \Lambda\Big(\rho_{BC}(\overline{\rho})\Big)
- \Lambda\Big(\rho_A(\overline{\rho})\Big).
\ee
In the ZRP case, one recovers $\alpha=\ln 2$, independently of the density
$\overline{\rho}$, while with continuous masses,
assuming $v(\mu)=1$ and $f(m)=m^{\eta-1}$, one finds
\be
\alpha(\overline{\rho}) = \frac{\eta}{\overline{\rho}}
\left( 1-2\kappa+\frac{3\kappa-1}{\sqrt{2}}\right).
\ee
This result is in agreement with the numerical findings of Sect.~\ref{sec-3branch-fluct}
where $\eta=1$ and $\kappa=\frac{1}{3}$, yielding $\alpha=1/(3\overline{\rho})$ (see right panel of Fig.\ref{fig-collapse}).
Hence, despite the absence of an exact solution
of the mass transport model (beyond the specific ZRP case),
this tentative scenario provides a consistent
explanation of the numerical results obtained for the local chemical potential,
showing that the asymptotic factorization condition may still hold while
local measurements fail to find an equilibrated chemical potential
throughout the system.

\section{Discussion and conclusion}
\label{sec-discuss}

In this paper, we have attempted to clarify the influence of the
flux balance condition on the generalized chemical potential
in non-equilibrium mass transport models.
We have seen in particular that even when the asymptotic
factorization condition holds,
locally measured values of the chemical potential may not equalize,
while the theoretically-defined ones remain equal.
This surprising property has been traced back to the presence
of exponential factors appearing in the probability weights of
different subsystems. These exponential factors cancel out
at the level of the local statistics (for instance the single-site
distribution) and are thus locally undetectable.
But on the other hand these factors play an essential
role in the global statistics to ensure the balance of flux.

Quite surpringly, such factors can exhibit a discontinuity between
equilibrium and weakly non-equilibrium situations, as exemplified
by the exact solution of the ZRP: at equilibrium (zero flux), no exponential
factor is present (and local chemical potentials equalize),
while the tiniest flux generates a factor $2^M$, leading to a shift
$\ln 2$ between the local chemical potentials.
We have also shown that a tentative generalization of this scenario,
assuming a density-dependent shift $\alpha(\overline{\rho})$,
is consistent with the numerical results.

An important consequence of these results is that in most cases it is not
possible to define along the present lines a locally measurable
chemical potential that would equalize throughout the system.
This comes from the fact that the shift $\alpha(\overline{\rho})$
depends on the global density of the system, which cannot be
measured locally.
Hence, although no long-range correlations are present,
a global information is required to relate what happens at two different
locations in the system.

\subsection*{Acknowledgements}

K.~M.~was supported by the Marie Curie FP7-PEOPLE-2009-IEF program.
Both authors thank M.~Droz for useful discussions at the early stage of this work.

\end{document}